%% file: 0000p4est3-avx.tex
\pdfoutput=1

\documentclass[twoside,leqno,twocolumn]{article}

\usepackage[letterpaper]{geometry}

\usepackage{ltexpprt}
\usepackage{hyperref}
\input{packages}

\input{macros}

\newtheorem{defn}{Definition}[section]
\newtheorem{proper}[defn]{Property}
\newtheorem{remark}[defn]{Remark}
\renewenvironment{Definition}{\begin{defn}}{\end{defn}}
\renewenvironment{property}{\begin{proper}}{\end{proper}}
\renewenvironment{Remark}{\begin{remark}}{\end{remark}}

\begin{document}

\newcommand\relatedversion{}


\title{\Large Alternative quadrant representations with Morton index and
AVX2 vectorization for AMR algorithms
within the \pforest software library\relatedversion}%
\author{Mikhail Kirilin\thanks{Institute for Numerical Simulation, University of Bonn}%
\and Carsten Burstedde\thanks{Institute for Numerical Simulation, University of Bonn}}%


\maketitle







\begin{abstract} \small\baselineskip=9pt We present a technical enhancement
within the \pforest software for parallel adaptive mesh refinement. In \pforest
mesh primitives are stored as octants in three and quadrants in two dimensions.
While, classically, they are encoded by the native approach using its spatial
coordinates and refinement level,
any other mathematically equivalent encoding might be used instead.
Recognizing this, we add two alternative representations to the classical,
explicit version, based on a long monotonic index and 128-bit AVX quad integers,
respectively. The first one requires changes in logic for low-level quadrant
manipulating algorithms, while the other exploits data level parallelism and
requires algorithms to be adapted to SIMD instructions. The resultant
algorithms and data structures lead to higher performance and lesser memory
usage in comparison with the standard baseline.
We benchmark selected algorithms on a cluster with two Intel(R) Xeon(R) Gold
6130 Skylake family CPUs per node, which provides support for AVX2
extensions, 192 GB RAM per node, and up to 512 computational cores in total.
\end{abstract}

\section{Introduction}
\label{sec:introduction}

The \pforest software \cite{BursteddeWilcoxGhattas11}
serves to create, refine and coarsen and to partition an adaptive mesh in
parallel, as well as to 2:1 balance the refinement pattern
\cite{IsaacBursteddeGhattas12}.
It is built around the concept of a forest of 2D quadtrees or 3D octrees,
providing quadrilateral and hexahedral elements, respectively.
General geometries are meshed by connecting multiple, logically cubic trees
into a forest, and optionally adding arbitrary geometry maps.
In addition, \pforest provides various algorithms to interrogate the mesh from
an application perspective.
These include a general ghost/halo layer construction, node numberings for
low- and high-order continuous elements, an interface iterator
\cite{IsaacBursteddeWilcoxEtAl15}, and flexible local and remote search
functions \cite{Burstedde20d}.
\pforest is one of the most scalable codes available, reaching $10^6$ MPI ranks
and more \cite{RudiMalossiIsaacEtAl15, MullerKoperaMarrasEtAl18}.

As the domain is logically refined as a collection of interconnected trees,
replacing selected nodes into $2^d$ child nodes recursively, a \pforest mesh
consists of the leaves only.
Ancestor nodes are constructed on demand and routinely used within top-down
traversal algorithms but not permanently remembered or referenced.
This principle has been introduced by Dendro \cite{SundarSampathBiros08} and
distinguishes \pforest from its historic predecessor octor
\cite{TuOHallaronGhattas05} and overlapping tree codes such as amrex, boxlib,
and chombo.
\pforest is often used indirectly through generic discretization and solver
libraries such as
\dealii \cite{BangerthBursteddeHeisterEtAl11}
and \petsc \cite{IsaacKnepley15}.


This paper introduces technical enhancements to \pforest that remain
invisible to most users but are measurable in terms of improved performance and
memory usage.
Specifically, we abstract the low-level representation of a quadrant to allow
for re-implementations with different speed and storage characteristics under a
virtual interface.
This first idea is not new in itself, cf.\ for example the upfront design of
the \tetcode \cite{BursteddeHolke16, HolkeBursteddeKnappEtAl23} that allows
for trees and elements shaped as $d$-cubes, triangles, tetrahedra, and more
recently prisms \cite{Knapp17} and pyramids \cite{Knapp20}.
What is new in this context
is the exact representations we add for quadrants and octants.
The first one is a raw space-filling curve index, to which a
similar idea was presented in \cite{KellerCavelanCabezonEtAl23} and used for an
octree forest construction. The differences are that the authors choose a Hilbert space
filling curve and present an algorithm for forest creation, while we
use the Morton curve and implement low-level bitwise operations to add
this representation to the entire AMR workflow. The
second representation is a hardware-oriented 128-bit field processed according
to Intel's Advanced Vector Extensions (AVX) 2.
These introduce data parallelism without
concurrency, when a unit executes the same instruction on multiple data
simultaneously at any given moment, which is its own type of parallel processing
\cite{Flynn72}.  In other contexts, the concept is
used to vectorize array loops or matrix operations; see e.g.\
\cite{ShabanovRybakovShumilin19}, or finite element solver loops
\cite{JubertieDuprosDeMartin18}. There are various approaches to realize
vectorization in a code rather than through compiler auto-vectorization: compiler
directives
as in
\cite{HadadeWangCarnevaleEtAl19}, intrinsics, which we present in this work,
and inline assembly.

We summerize the following list of contributions:
\begin{enumerate}
    \item Add support for quadrants encoded into a monotonic raw Morton
    index.  The use of this implementation reduces the computational
    complexity for some low-level algorithms.
    \item Add support for quadrants loaded into extended 128-bit CPU
    registers and adapt internal algorithms to utilize SIMD
    instructions implemented via AVX2.
    This allows to increase the maximum possible refinement level and
    boosts performance.
    \item Compare differences in low-level algorithms design, logic, and
    complexity depending on the quadrant representation used.
    \item We present a series of synthetic tests designed to evaluate the
    runtime speedups achieved by the updated algorithms in combination with
    new quadrant implementations on Skylake CPU nodes of the Bonna cluster
    at the University of Bonn.
    \item Demonstrate the impact of our manual
    vectorization on
    performance in comparison to the builtin compiler vectorization.
    \item Benchmark the consumption of RAM by our implementations and
    compare them to each other as well as to the standard representation. Our
    implementation loaded into 128-bit SIMD registers reduces RAM usage by the
    factor of 1/3 and the raw Morton index implementation by 2/3.
\end{enumerate}

Through our technical enhancements and the subsequent analysis and
evaluations, we aim to contribute to the advancement of adaptive mesh
refinement techniques in general.
Specifically, we improve the overall performance and memory
consumption of the \pforest software.
These improvements constitute only the first part of a larger development
to be presented in follow-up papers:
\begin{enumerate}
    \item Using updated low-level algorithms, we introduce high-level ones
    that utilize new quadrant implementations and bring new functionality
    to the \pforest workflow.
    \item We integrate MPI-3 shared memory support to take advantage
    of the architecture of single shared memory nodes, reduce the number of
    messages (i.e.\ providing communication-free partition and mesh
    neighbor iteration) and decrease the amount of RAM usage by replicated data.
    \item Remove obsolete bits from quadrant coordinates, previously used
    to shift outside the unit tree and to designate mesh nodes, to increase the
    attainable maximum refinement by 3 levels.
    \item Present a mesh iteration algorithm that is functional in the
    presence of non-2:1-balanced meshes. Previously, we had been requiring
    balance to traverse all interfaces between quadrants and the boundary.
\end{enumerate}

We make the code for our developments available inside the public \pforest
repository \cite{Burstedde23a} (for the time being as a compilable copy of
the relevant files, not the exact history, which will be rectified when the
remaining developments are published).

\section{Linear octree storage and representation}
\label{sec:linearstorage}

We have recently summarized the data stored in a \pforest object to uniquely
define the mesh \cite{Burstedde20d}, which we refer to for details.
Here we will only introduce the notation absolutely necessary:
The principal parameters for a forest are the number of trees
$K$, the global number of quadrants $N$ (we continue to use the word quadrants
even in 3D), and the number of MPI processes $P$.
The quadrants form a disjoint union of all leaves in the forest and are
partitioned between the MPI processes in the order of a space filling curve.
This curve is currently the well-known Morton or Z-curve \cite{Morton66,
    TropfHerzog81}, which we allow to be replaced as long as all required
operations on and between quadrants are provided.

The current (standard) representation of a quadrant contains the coordinates
of its (lower) front left corner and its refinement level.
This information is sufficient to compute its ancestor and first and last
descendants on any level, in particular its parent, and any child and sibling.
We may construct neighbors across any given face, edge, or corner.
We may also interrogate the quadrant for its child number relative to a parent
or higher ancestor.
A number of typical per-quadrant operations, all of which execute in $\cO(1)$
time, are printed in the original paper on \pforest
\cite{BursteddeWilcoxGhattas11}, while we refer to the source code for all
others \cite[\texttt{src/p4est\_bits.c}]{Burstedde23a}.

In addition to modifications of quadrants, we may want to translate a
quadrant's location and length into a space filling curve index and back.
These operations tend to overflow the integer range when not enough bits
are provided.
In practice, the original \pforest software packs either 2D or 3D coordinates
into 64 bits, which together with added bits for encoding neighbors outside the
unit tree sets the maximum refinement level of a quadrant to 29 and an octant
to 18, respectively.
We describe elsewhere how we raise the 3D limit to the same 29 using 128-bit
indices.
In this paper, on the other hand, we encounter different maximum levels
depending on the internal representation of quadrants.
Perspectively, we aim to minimize the use of the curve index since it is
rarely ever needed.

Since the inception of \pforest, we work with self-sufficient quadrant data,
meaning that each  quadrant object contains full information on coordinates
and level and thus space filling curve position.
This allows for random access into sets of quadrant arrays and the generation
of temporary quadrants in top-down traversals and searches.
An alternative approach is to run-length encode a range of quadrants using the
mathematical logic of the space filling curve \cite{BaderBockSchwaigerEtAl10,
    WeinzierlMehl11}.
This reduces storage size further (which is already small compared to numerical
data) but removes the flexibility to access or iterate out of order, vertically
(i.e., changing the level) or horizontally (changing coordinates).

While a hardcoded, fixed quadrant representation is the most direct,
it lacks the flexibility to experiment and optimize further.
To this end, we abstract the quadrants' implementation to be varied while their
logical information remains equivalent.
We follow two goals, namely to allow for different space filling curves and
orderings while writing the octree algorithms just once, and to allow for
efficient hardware-oriented implementations of quadrant-level functions.
This paper focuses on the second aspect, while the first is reserved for future
research.

We have made explicit the separation of meshing software into high- and
low-level algorithms in the \tetcode papers \cite{BursteddeHolke16,
    HolkeKnappBurstedde21}.
Low-level algorithms are the per-quadrant operations listed just above,
and high-level algorithms are for example the initial forest construction,
refinement, and partitioning, which in turn rely on low-level operations
arranged in forest traversals and loops.
The idea is to change between multiple sets of quadrant representations and
associated low-level operations using the same high-level algorithm.

\subsection{Standard representation: $xyz$ and level.}
\label{sec:stand}

First of all, we note that the coordinates of a quadrant within a unit tree
cube are integer multiples of the quadrant's length, which is halved for each
additional refinement level beginning with the root at length 1.
Thus, to express the coordinates in binary, a fixed point representation would
be ideally suited, while a floating point number wastes the exponent bits.
The third option, and a rather practical one, is to limit the refinement to a
given maximum level $L$ and to express coordinates and length in integer space.
This effectively leads to the ranges $x, y, z \in \halfopen{0, 2^L} \cap \sZ$
and an integer quadrant length at level $\ell$ of $h = 2^{L - \ell}$.

Given the refinement level $\ell$ of a quadrant, its index relative to the
space filling curve may either be relative to its own level, i.e.\ $I_\ell \in
\halfopen{0, 2^{d\ell}}$, or to the maximum level $L$, with $0 \le \ell  \le L$
and $I = I_L \in \halfopen{0, 2^{dL}}$.
Both transform into each other by bit shifts.
Our choice is to work relative to the maximum level, since adaptive meshes
contain quadrants of mixed levels and we eliminate shift operations when
creating ancestors or descendants.
As is well known, the index of a quadrant with respect to the Morton space
filling curve relative to the maximum level $L$ is obtained by the bitwise
interleaving of coordinates.

\begin{algorithm}
    \caption{\\ \fxn{Standard\_Morton}(\vartype{uint64} $I_{\ell}$,
    \vartype{uint8} $\ell$) $\rightarrow q$}
    \footnotesize
    \label{alg:standard_morton}%
    \KwIn{Quadrant index $I_{\ell}$ wrt.\ level $\ell$ uniform mesh.}
    \KwResult{Quadrant $q = (xyz, \ell)$  with given
        index $I_\ell$ in standard representation.}
    \algrule
    $x \leftarrow 0,\ y \leftarrow 0,\ z \leftarrow 0$ \Comm*[f]{initialize
    output variables}\\
    \For{$0 \leq i < \ell $}
    {
        $\fxn{extract}_\mathrm{id} \leftarrow 1 \ShiftLeft d i$
        \Comm*[f]{mask to extract coordinate bit}\\
        $\fxn{shift}_\mathrm{crd} \leftarrow (d-1) i$ \Comm*[f]{shift
            to place the extracted bit}\\
        $x\ \bitwor\hspace{-1mm}= (I_{\ell} \bitwand
        (\fxn{extract}_\mathrm{id} + 0)) \ShiftRight (\fxn{shift}_\mathrm{crd}
        + 0)$\\
        $y\ \bitwor\hspace{-1mm}= (I_{\ell} \bitwand
        (\fxn{extract}_\mathrm{id} + 1)) \ShiftRight (\fxn{shift}_\mathrm{crd}
        + 1)$\\
        $z\ \bitwor\hspace{-1mm}= (I_{\ell} \bitwand
        (\fxn{extract}_\mathrm{id} + 2))
        \ShiftRight (\fxn{shift}_\mathrm{crd} + 2)$
    }
    $x\ \leftarrow x \ShiftLeft (L - \ell), \ldots$
    \Comm*[f]{set $x$, $y$, $z$ according to $L$}
\end{algorithm}%

\begin{algorithm}
    \caption{\\ \fxn{Standard\_Child} (\vartype{quad} $q$,
    \vartype{int} $c$)
        $\rightarrow \hat q$}
    \footnotesize
    \label{alg:standard_child}%
    \KwIn{Quadrant $q = (xyz,\ell)$, $\ell < L$. Index $c \in [0,\ldots,2^{d})$
    of
    a child to be constructed.}
    \KwResult{Quadrant $\hat q = (\hat{x}\hat{y}\hat{z}, \hat{\ell})$ in
    standard representation, the $c$-th child of $q$.}
    \algrule
    $\fxn{shift} \leftarrow 1 \ShiftLeft (L - (\ell + 1))$
    \Comm*[f]{shifting of a child}\\
    $\hat{x} \leftarrow \ternary{c \bitwand 001_2}{x \bitwor
        \fxn{shift}}{x}$ \Comm*[f]{$c$ indicates shifting\ldots}\\
    $\hat{y} \leftarrow \ternary{c \bitwand 010_2}{y \bitwor
        \fxn{shift}}{y}$ \Comm*[f]{\ldots to specific direction}\\
    $\hat{z} \leftarrow \ternary{c \bitwand 100_2}{z \bitwor
        \fxn{shift}}{z}$ \Comm*[f]{\ldots to add coordinate bit}\\
    $\hat{\ell} \leftarrow \ell + 1$
\end{algorithm}%

\begin{algorithm}
    \caption{\\ \fxn{Standard\_Sibling}
        (\vartype{quad} $q$, \vartype{int} $s$) $\rightarrow \hat q$}
    \footnotesize
    \label{alg:standard_sibling}%
    \KwIn{Quadrant $q  = (xyz,\ell)$, $\ell > 0$. Index $s \in
    [0,\ldots,2^{d})$ of
    a sibling to be constructed.}
    \KwResult{Quadrant $\hat q = (\hat{x}\hat{y}\hat{z}, \ell)$ in standard
    representation, the $s$-th sibling of $q$.}
    \algrule
    $\fxn{shift} \leftarrow 1 \ShiftLeft (L - \ell)$ \Comm*[f]{shifting of a
    sibling}\\
    $\hat{x} \leftarrow \ternary{s \bitwand 001_2}{x \bitwor
        \fxn{shift}}{x \bitwand \bitwneg \fxn{shift}}$
     \Comm*[f]{add shift\ldots}\\
    $\hat{y} \leftarrow \ternary{s \bitwand 010_2}{y \bitwor
        \fxn{shift}}{y \bitwand \bitwneg \fxn{shift}}$
      \Comm*[f]{or blank\ldots}\\
    $\hat{z} \leftarrow \ternary{s \bitwand 100_2}{z \bitwor
        \fxn{shift}}{z \bitwand \bitwneg \fxn{shift}}$
    \Comm*[f]{as necessary}
\end{algorithm}%

The maximum level of the $(xyz, \ell)$ representation then has two limits: a
larger by the number of bits, presently 32, of each coordinate, and an often
smaller one determined by the bit length chosen for $I$, presently 64.
An example of the latter kind is the construction of a standard quadrant from
its Morton index  (\algref{standard_morton}).  While this algorithm
is first published in \cite{BursteddeWilcoxGhattas11} we list it here as a
reference to compare how it changes with various quadrant implementations.

For performance reasons, it is generally desirable to eliminate the use of
the index $I$ and to rely on quadrant-relative operations where possible.
This is the case for example for high-level algorithms for refinement, which
can iteratively reach the maximum level without calling the Morton
transformation.
Here we use transformations as in
%
\algref{standard_child} and
\algref{standard_sibling}, which derive a child and a sibling of a given
quadrant by modifying coordinate bits.

Mathematically, we rely on the following facts.
\begin{Definition}
        Quadrant $r$ is the \emph{$c$-th child} of quadrant $q = (xyz,
        \ell)$, $\ell < L$ if and only if
	\begin{itemize}
		\item[1)] its level is equal to $\ell + 1$,
		\item[2)] its space filling curve related index $I_{\ell + 1}$ is
		derived from $q$'s related index as follows:
  \begin{equation*}
                I_{\ell + 1} = 2^d \times I_{\ell} + c ,
                \quad c \in [0, \ldots, 2^d) .
  \end{equation*}
	\end{itemize}
\end{Definition}

\begin{property}
        Each quadrant is the child of precisely one quadrant, its parent.
        Every 0-th child has the same coordinates $xyz$ as its parent
        but half its size.
	The $c$-th child $r$ is a half-size quadrant located inside its
	parent quadrant $q$ and positioned relative to the 0-th child
        according to $d$ direction bits stored in $c$.
\end{property}
According to this logic, \algref{standard_child}
creates a quadrant's child setting up to $d$ bits,
without using any space filling curve indices.


\begin{Definition}
        Quadrant $r$ is the \emph{$s$-th sibling} of quadrant $q = (xyz,
        \ell)$, $\ell > 0$ if and only if
	\begin{itemize}
		\item[1)] its level is equal to $\ell$,
		\item[2)] its space filling curve index $\hat{I}_{\ell}$ is
		derived from $q$'s index as follows:
  \begin{equation*}
                \hat{I}_{\ell} = I_{\ell} - (I_{\ell} \bmod 2^d) + s ,
                \quad s \in [0, \ldots, 2^d) .
  \end{equation*}
	\end{itemize}
\end{Definition}

\begin{property}
	A sibling $r$ to the quadrant $q$ is of the same size and moved
	along the space filling curve forward or backward relative to $q$.
        There is precisely one quadrant $p$ at level $\ell - 1$ such that $q$
	and $r$ are both its children and $r$'s child index is $s$.
\end{property}
Analogously to \algref{standard_child}, \algref{standard_sibling} creates a
sibling without use of $q$'s curve index $I_{\ell}$.

All three algorithms operate on a quadrant in $(xyz, \ell)$ representation.
Historically, this standard representation includes eight bytes of user
data/payload for a quadrant size of 16 bytes in 2D and 24 bytes in 3D.

\subsection{New representation: raw Morton index.}
\label{sec:mort}

A rather idiosyncratic representation of a quadrant is the Morton index $I$
itself.
This approach can yield an advantage in storage size if a conservative number
of bits is used for $I$.
However, we like to encode the level along with the index, which can be
realized in practice by storing the level in the 8 high bits of a 64-bit
integer and the index in the low 56 bits.
(We might turn this order around but see no apparent advantage to it.)
Since we define a new representation without regard to legacy algorithms, we
use all 56 bits for coordinates inside the unit tree, which leads to a maximum
level of $18 = \floors{56 / 3}$ in 3D.
This is the same as in original \pforest.

The main advantage of the raw Morton representation is its minimized storage
size and that the transformation to the Morton index is the identity, i.e.\
\algref{morton_morton} corrects the quadrant's level-specific index
$I_{\ell}$ to turn it into the level-independent $I$.
The successor operation, which, as is shown in Algorithm
\ref{alg:morton_successor}, derives the subsequent quadrant following a Morton
curve, reduces to a summation $I \leftarrow I + (1 \ShiftLeft d(L - \ell))$.
Meanwhile, the current level $\ell$ is accessed from $I$ by right shifting it
by 56.
The main disadvantage is that several other operations, such as Parent and
Child, become slightly less straightforward compared to the standard
representation; see e.g.\ Algorithm \ref{alg:morton_child} and Algorithm
\ref{alg:morton_parent}.
\begin{Definition}
    \label{defn:parent}
	Quadrant $r$ is a \emph{parent} of quadrant $q = (xyz, \ell),\ \ell > 0$
	if and only if
	\begin{itemize}
		\item[1)] its level is equal to $\ell - 1$,
		\item[2)] its space filling curve related index $I_{\ell - 1}$ is
		derived from $q$'s related index as follows: $$I_{\ell - 1} =
		\frac{I_{\ell} - (I_{\ell} \bmod 2^d)}{2^d}.$$
	\end{itemize}
\end{Definition}
%
%
\algref{morton_parent} creates a parent $r$ of the quadrant $q = (I, \ell)$.
According to Definition \ref{defn:parent}, it takes the index bits
responsible for the positioning on level $\ell$ and makes them zero or, in
other words, identical to $r$'s 0-th child.

Furthermore, let us present Algorithm \ref{alg:face_neigh} to create a
quadrant's neighbor across a face.
\begin{Definition}
    Consider a standard quadrant $q = (x_0 \ldots x_{d-1}, \ell)$.
    A d-dimensional cube
    $$
        \textup{dom}(q) := \big\{X \in \mathbb{R}^d\ :\ x_i \leq X \le x_i +
        2^{L - \ell}, 0 \leq i < d \big\}
    $$
    with side of length $2^{L - \ell}$ is called a quadrant
    $q$'s \emph{domain}.
\end{Definition}
\begin{Definition}
    Two quadrants $q$ and $r$ are \emph{face neighbors} if and only if
   	\begin{itemize}
        \item[1)] their levels are equal,
        \item[2)] the topological dimension of intersection of their domains is
        one less than d:
        $$
            \textup{dim} (\textup{dom} (q) \cap \textup{dom} (r)) = d-1.
        $$
    \end{itemize}

\end{Definition}
%
The integer $i \in [0,\ldots,2d)$
denotes the quadrant's face ordered as follows: faces across $x$ before
$y$ before, where applicable, $z$. Considering a level related Morton index as
a bitwise interleaving of coordinates, the quadrant $q$ is
Morton-represented as the sequence of bits
$$q = \overbrace{l_0 \ldots l_7}^{\text{level}}\ 00\
\overbrace{q_1^z q_1^y q_1^x\ q_2^z q_2^y q_2^x \ldots q_{18}^z q_{18}^y
q_{18}^x}^{\text{coordinates}}.$$
\begin{Remark}
    All bits after (right of) $q_{\ell}^x$ are equal to 0.
\end{Remark}
We construct a face neighbor with the following steps.
First we form an auxiliary $\fxn{mask}_{\mathit{dir}}$ that holds
the following bits:
$$\fxn{mask}_{\mathit{dir}} = \overbrace{0\ldots0}^{10}\ m_1^z m_1^y
m_1^x\ldots m_{\ell}^z m_{\ell}^y m_{\ell}^x\ 0\ldots0,$$ where in each group of
three $m_j^z m_j^y m_j^x,\ j \in \{1,\ldots,18\}$ only the bit responsible for
the target face neighbor direction is not zero. For consistency, let us assume
that we construct a neighbor across a face in $y$ direction. The the mask turns
into: $$\fxn{mask}_{\mathit{dir}} = \overbrace{0\ldots0}^{10}\ \overbrace {0 1
0\ldots0 1 0}^{\ell}\ 0\ldots0,$$ where masks responsible for $x$ and $z$ may
be derived from it by one bit shift right or left, respectively.
If the target neighbor is along the coordinate axis, we apply the negation of
the $\fxn{mask}_{\mathit{dir}}$ to $q$, preserving all 1-bits
except the ones that define the axis direction:
$$q \bitwor \bitwneg \fxn{mask}_{\mathit{dir}} = \overbrace{1\ldots1}^{10}\ 1
q_1^y 1\ldots1 q_{\ell}^y 1\ 1\ldots1.$$
Incrementing this value by 1 means increasing $q$'s $y = q_1^y \ldots
q_{\ell}^y\ 0\ldots0$ coordinate by $2^{\ell}$, the quadrant length.

Analogously, we shift the quadrant against the $y$ axis, but using the conjunction
operation, non-negated $\fxn{mask}_{\mathit{dir}}$ and decrementing by 1.
Finally we must restore the bits that stay unchanged; see
Algorithm \ref{alg:face_neigh}.
%

\begin{algorithm}
    \caption{\\ \fxn{Morton\_Morton}
        (\vartype{uint64} $I_{\ell}$, \vartype{uint8} $\ell$) $\rightarrow q$
    }
    \footnotesize
    \label{alg:morton_morton}%
    \KwIn{Quadrant index $I_{\ell}$ wrt.\ level $\ell$ uniform mesh.}
    \KwResult{Quadrant $q = (I, \ell)$ with given level-specific
        index $I_\ell$ in raw Morton representation.}
    \algrule
    $q \leftarrow \ell \ShiftLeft 56$ \\
    $q \;\bitwor\hspace{-1mm}= I_{\ell} \ShiftLeft d (L - \ell)$
    \Comm*[f]{relate the index to maximum level $L$}
\end{algorithm}%

\begin{algorithm}
    \caption{\\ \fxn{Morton\_Successor}
        (\vartype{uint64} $q$) $\rightarrow r$}
    \footnotesize
    \label{alg:morton_successor}%
    \KwIn{Quadrant $q = (I, \ell)$ in raw Morton representation; not the last
    one.}
    \KwResult{Raw Morton quadrant $r$ with the level-specific index
    $\hat{I_\ell} = I_{\ell} + 1 < 2^{d\ell}$.}
    \algrule
    $r \leftarrow q + (1 \ShiftLeft d (L - (q \ShiftRight 56)))$
    \Comm*[f]{increase by 1 wrt.\ $L$}
\end{algorithm}%

\begin{algorithm}
    \caption{\\ \fxn{Morton\_Child}
        (\vartype{uint64} $q$, \vartype{int} $c$) $\rightarrow r$}
    \footnotesize
    \label{alg:morton_child}%
    \KwIn{Morton quadrant $q = (I,\ell)$,\ $\ell < L$. Index $c \in
    [0,\ldots,2^{d})$ of a child to be construced.}
    \KwResult{Quadrant $r = (\hat{I}, \hat{\ell})$ in
              raw Morton representation, the $c$-th child of $q$.}
    \algrule
    $\fxn{shift} \leftarrow c \ShiftLeft d (L - ((q \ShiftRight 56) +
    1))$ \Comm*[f]{shifting of a child}\\
    $r \leftarrow (q \bitwor \fxn{shift})
 + (1 \ShiftLeft 56)$
\end{algorithm}%

\begin{algorithm}
    \caption{\\ \fxn{Morton\_Parent}
        (\vartype{uint64} $q$) $\rightarrow r$}
    \label{alg:morton_parent}%
    \footnotesize
    \KwIn{Morton quadrant $q = (I,\ell)$,\ $\ell > 0$.}
    \KwResult{Quadrant $r = (\hat{I}, \hat{\ell})$ in
              raw Morton representation, the parent of $q$.}
    \algrule
    $r \leftarrow q \bitwand \bitwneg (111_2 \ShiftLeft d (L - (q \ShiftRight
    56)))$ \Comm*[f]{blank bit on\ldots}\\
    $r \leftarrow r - (1 \ShiftLeft 56)$
    \Comm*[f]{\ldots level $\ell$ responsible bits}
\end{algorithm}%

\begin{algorithm}
    \caption{\\ \fxn{Morton\_FNeigh}
        (\vartype{uint64} $q$,\ \vartype{int} $i$) $\rightarrow r$
    }
    \footnotesize
    \label{alg:face_neigh}%
    \KwIn{Quadrant $q = (I,\ell)$. Face index $i \in [0,\ldots,2d)$, along
    which the neighbor to be constructed.}
    \KwResult{Raw Morton quadrant $r = (\hat{I}, \ell)$, the neighbor of
    $q$ across $q$'s $i$-th face.}
    \algrule
    $\fxn{sign} \leftarrow \ternary{i \bitwand 1}{1}{-1}$\\
    $\fxn{mask}_l \leftarrow \bitwneg((1 \ShiftLeft d (L - (q \ShiftRight 56)))
    - 1)$\\
    $\fxn{mask}_{\mathit{dir}} \leftarrow (\overbrace{0\ldots0}^{10}
    \overbrace{001001\ldots001_2}^{54} \bitwand \fxn{mask}_l) \ShiftLeft
    \floors{\frac{i}{2}}$\\
    \eIf{$\fxn{sign} = 1$}{
        $r \leftarrow (q \bitwor \bitwneg \fxn{mask}_{\mathit{dir}}) + 1$
        \Comm*[f]{move along axis direction} \\
    }
    {
        $r \leftarrow (q \bitwand \fxn{mask}_{\mathit{dir}}) - 1$
        \Comm*[f]{move against axis direction} \\
    }
    $r \leftarrow (r \bitwand \fxn{mask}_{\mathit{dir}}) \bitwor (q
    \bitwand \bitwneg \fxn{mask}_{\mathit{dir}})$
    \Comm*[f]{restore changed bits}
\end{algorithm}%

\subsection{New representation: 128-bit SIMD/AVX2.}
\label{sec:AVX}

Given that an octant is defined by four values $x$, $y$, $z$ and $\ell$, it
seems natural to consider four-way SIMD (Single Instruction Multiple Data)
instructions for accelerated processing.
While there is a certain asymmetry between the coordinates and the level,
packing the manipulation of all coordinates into one instruction is
nevertheless an intriguing possibility.

To implement this idea, we base our second new quadrant representation on the
Advanced Vector Extensions 2 and legacy Streaming SIMD Extensions (SSE/AVX2).
SSE/AVX2 is a set of the SIMD
processor intrinsics developed by Intel and widely supported in CPU hardware.
These intrinsics operate on extended processor registers that contain wider
data than the regular ones.
In our example, we use AVX2 to operate on four 32-bit numbers at a time.
%
Specifically, we chose the
special SSE2 type \mitype.
\begin{figure}
    \begin{minipage}[b]{\columnwidth}
        \begin{equation*}
            \text{Quadrant} :=
            \begin{cases}
                x & = 0 \; x_0 \; x_1 \; x_2 \; \ldots \; x_{30}\\
                y & = 0 \; y_0 \; y_1 \; y_2 \; \ldots \; y_{30}\\
                z & = 0 \; z_0 \; z_1 \; z_2 \; \ldots \; z_{30}\\
                \text{level} & = 0 \; \ldots \; 0 \; l_0 \; l_1 \;
                \ldots \; l_7
            \end{cases}
        \end{equation*}
    \end{minipage}%
    \begin{center}
        \includegraphics[width=\columnwidth]{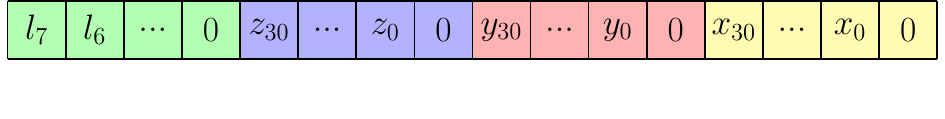}
    \end{center}
    \caption{One way of storing a quadrant's bit representation (above) in an
        extended SIMD 128-bit register (below).}\label{fig:avx}%
\end{figure}%
It stores 128 bits of data interpreted as signed integers; see Figure
\ref{fig:avx}. The data are stored in reverse order due to peculiarities of
storage and processing information in extended registers.
This approach to quadrant data storage natively resolves the problem of the
data alignment required by a SIMD unit.

Converting \pforest's low-level quadrant operations to AVX2 requires not only
the Intel's intrinsics use, but also modifications for the algorithms' design.
%
%
Thereby we reduce the number of mathematical operations in most of the
algorithms.
As an example, consider the listings in Algorithm \ref{alg:avx_child} and
Algorithm \ref{alg:avx_parent}.
\begin{algorithm}
    \caption{\\ \fxn{AVX\_Child}
        ($\vartype{quad}_{\vartype{AVX}}\ q$, \vartype{int} $\mathit{c}$)
        $\rightarrow r$}
    \footnotesize
    \label{alg:avx_child}%
    \KwIn{Quadrant $q = \AVXArray{\ell}{z}{y}{x}$,\ $\ell < L$. Index $c
    \in [0,\ldots,2^{d})$ of a child to be construced.}
    \KwResult{Quadrant $r = \AVXArray{\hat{\ell}}{\hat{z}}{\hat{y}}{\hat{x}}$
    that is the $c$-th child of $q$ packed into 128-bit SIMD register.}
    \algrule
    \ali{4em}{$\fxn{ext}_{\mathit{id}}$} $\xleftarrow[]{128}
    \AVXArray{001^{32}_2}{010^{32}_2}{100^{32}_2}{000^{32}_2}$\\

    \ali{4em}{$\fxn{ext}_{\mathit{id}}$} $\xleftarrow[]{128}
    \fxn{ext}_{\mathit{id}} \bitwand
    \AVXArray{\mathit{c}}{\mathit{c}}{\mathit{c}}{0}$\\

    \ali{4em}{$\fxn{ins}_{\mathit{id}}$} $\xleftarrow[]{128}
    \fxn{ext}_{\mathit{id}} \ShiftRight_{32}
    \AVXArray{0^{32}}{1^{32}}{2^{32}}{0^{32}}$\\

    \ali{4em}{$r$} $\xleftarrow[]{128} q \bitwor
    (\fxn{ins}_{\mathit{id}} \ShiftLeft (L - (\ell + 1)))$\\
    \ali{4em}{$r[0] = \hat{\ell}$} $\xleftarrow[]{\;\;\;\;\;} \ell + 1$
    \Comm*[f]{insert value $\ell + 1$ into 0-th position}
\end{algorithm}%
%
\begin{algorithm}
    \caption{\\ \fxn{AVX\_Parent}
        ($\vartype{quad}_{\vartype{AVX}}\ q$) $\rightarrow r$}
    \footnotesize
    \label{alg:avx_parent}%
    \KwIn{128-bit quadrant $q = \AVXArray{\ell}{z}{y}{x}$,\ $\ell > 0$.}
    \KwResult{Quadrant $r = \AVXArray{\hat{\ell}}{\hat{z}}{\hat{y}}{\hat{x}}$
    packed into 128-bit SIMD register, the parent of $q$.}
    \algrule
    \ali{3.5em}{$\fxn{len}$}$\xleftarrow[]{\;\;\;\;\;} 1 \ShiftLeft (L -
    \ell)$\\
    \ali{3.5em}{$r$}$\xleftarrow[]{128} q \bitwand
    \AVXArray{\neg\fxn{len}}{\neg\fxn{len}}{\neg\fxn{len}}{\neg0}$\\
    \Comm*[f]{blank child's coordinates bits}\\
    \ali{3.5em}{$r[0] = \hat{\ell}$}$\xleftarrow[]{\;\;\;\;\;} \ell - 1$
\end{algorithm}%
%
The block array $\AVXArray{n}{..}{..}{1}$ denotes an
extended register that is treated as $n$ 32-bit integer numbers.
We tag some values with a subscript number $s$ to emphasize their capacity
and mark some non-obvious assignment operators ,,$\xleftarrow[]{m}$`` with the
value $m$, which specifies the number of bits copied to a register.
For consistency and due to the structure of SSE/AVX2 registers we fix $n
\times s = m$.
Binary operators are divided into two groups: the first treats
bits in a register as a whole number, which applies to the regular
non-extended registers as well; the others operate on the data in a register
as a sequence of $n$ $s$-bit numbers.
We denote the first group by the usual operator symbol ,,\fxn{op}``, while the
other group with subscript ,,$\fxn{op}_s$`` indicates the capacity of a
register's subdivision, where
 $\fxn{op} \in \{\ShiftLeft,\ \ShiftRight,\ \&,\ \ -,\ == \}$.
Thus, the vectorized version
of the $\fxn{Child}$ function, presented originally in Algorithm
\ref{alg:standard_child}, takes 30--46\% less mathematical operations (7
vs.\ 10--13 depending on conditional results).

Although we choose \mitype to store four 32-bit as a main type,
we need even more
bits
in case of overflowing
due to some operations.
An example for this case is listed in
Algorithm \ref{alg:avx_morton}
for creating a new quadrant by its Morton index and level.
Since the Morton index $I_{\ell}$ occupies 64 bits and similarly
to Algorithm \ref{alg:standard_morton}, it needs temporarily $d \times L = 3
\times 18 = 54$ bits, consequently a 64-bit variable is needed to store
and process each single coordinate.
To solve this problem we stay with the 128-bit SIMD registers but
restrict ourselves to processing only two coordinates at the same time
and calculating the third one separately.
A second approach, to process all tree coordinates simultaneously,
would require us to temporarily use 256-bit (SSE/AVX2) registers provided by
the special AVX2 type \mitypeex.
According to our experiments (not shown here), mixing register
lengths
leads to a significant
slowdown, even though the task appears to be parallelized better.



We also demonstrate the SSE/AVX implementation of Algorithm
\ref{alg:avx_tree_bound} for finding the face numbers
of a tree that are touched by a given quadrant
\cite{IsaacBursteddeWilcoxEtAl15}.
As a result, it returns an array $f$ of indices of tree faces that
intersect the quadrant. The array's length is $d$ and its values depend on
which boundary face is touched along the $i$-th direction:
\begin{equation*}
    f[i] :=
    \begin{cases}
        -2, & \textit{all boundaries (i.e.\ $\ell = 0$)}\\
        -1, & \textit {no boundary}\\
        \phantom{-}2 \times i, & \textit{first face along $i$-th direction}\\
        \phantom{-}2 \times i + 1, & \textit{second face along $i$-th direction}
    \end{cases}
\end{equation*}
After processing the
trivial case $\ell = 0$, we check if the quadrant touches any boundary and
then place a necessary value in the corresponding location of the register.
Initially, we assign the array with the values one more than we want to
return and, consequently, subtract 1 so as to distinguish the case when the
quadrant does not intersect the boundary from intersection the first face
along the $x$-axis.

\begin{algorithm}
    \caption{\\ \fxn{AVX\_Morton}
        (\vartype{uint64} $I_{\ell}$, \vartype{uint8} $\ell$) $\rightarrow q$}
    \footnotesize
    \label{alg:avx_morton}%
    \KwIn{Quadrant index $I_{\ell}$ wrt.\ level $\ell$ uniform mesh.}
    \KwResult{Quadrant $q = \AVXArray{\ell}{z}{y}{x}$ with given index $I_\ell$
    packed into 128-bit register.}
    \algrule
    $q = \AVXArray{\ell^{32}}{z^{32}}{y^{32}}{x^{32}}
    \xleftarrow[]{128} \AVXArray{0}{0}{0}{0}$\\
    \Comm*[f]{initialize 128-bit SIMD register}\\
    \For{$0 \leq i < \ell $}
    {
        {$\fxn{x}_{\mathit{id}}$} $\leftarrow d i$,\
        $\fxn{x}_{\mathit{crd}}$ $\leftarrow (d - 1) i$\\
        \ali{3em}{$\fxn{ext}_{\mathit{id}}$} $\xleftarrow[]{128}
        \AVXArrayRed{1^{64}}{1^{64}} \ShiftLeft_{64}
        \AVXArrayRed{\fxn{x}_{\mathit{id}} +
            1}{\mathit{\;\;\;\;}\fxn{x}_{\mathit{id}}\;\;\;\;}$\\
        \ali{3em}{$\fxn{crd}_{\mathit{id}}$} $ \xleftarrow[]{128}
        \AVXArrayRed{\mathit{\;I_{\ell}\;}}{\mathit{\;I_{\ell}\;}} \bitwand
        \fxn{ext}_{id}$\\
        \ali{3em}{$\fxn{crd}_{\mathit{id}}$} $ \xleftarrow[]{128}
        \fxn{crd}_{\mathit{id}} \ShiftRight_{64}
        \AVXArrayRed{\fxn{x}_{\mathit{crd}}
            + 1}{\mathit{\;\;\;\;}\fxn{x}_{\mathit{crd}}\;\;\;\;}$\\
        \ali{3em}{$\fxn{crd\_2}_{\mathit{id}}$} $
        \xleftarrow[]{\;\;\;\;\;}
        (I_{\ell} \bitwand (1^{64} \ShiftLeft (\fxn{x}_{\mathit{crd}} + 2)))
        \ShiftRight (\fxn{x}_{\mathit{crd}} + 2)$\\
        \ali{3em}{$q$} $ \xleftarrow[]{128} q \bitwor \AVXArray
        {0}{\fxn{crd\_2}_{\mathit{id}}}{\fxn{crd}_{\mathit{id}}[0]}
        {\fxn{crd\_2}_{\mathit{id}}[1]}$
    }
    \ali{1.7em}{$q$} $\xleftarrow[]{128} q \ShiftLeft (L-\ell)$\\
    \Comm*[f]{shift each of four integers left to set coordinates wrt.\ L}\\
    \ali{1.7em}{$q[0]$} $\xleftarrow[]{\;\;\;\;\;} {(\vartype{uint32}) \ \ell}$
\end{algorithm}%

\begin{algorithm}
    \caption{\\ \fxn{AVX\_Tree\_Boundaries}
        ($\vartype{quad}_{\vartype{AVX}}\ q$) $\rightarrow f[d]$}
    \footnotesize
    \label{alg:avx_tree_bound}%
    \KwIn{Quadrant $q = \AVXArray{\ell}{z}{y}{x}$, that will be tested on
    touching three's boundary.}
    \KwResult{Array $f$ of length $d$ storing face indices.}
    \algrule
    \If{$\ell = 0$}{
        $f[0] \leftarrow f[1] \leftarrow f[2] \leftarrow -2$\\
        \textbf{return}
    }
    \ali{1.5em}{$\mathit{up}$}$\xleftarrow[]{\;\;\;\;\;} (1 \ShiftLeft L) -
    (1 \ShiftLeft (L - 1))$\\
    \Comm*[f]{Maximum length where a quadrant might locate}\\

    \If{$q =_{32} \AVXArray{\;0\;}{\ 0\;}{\ 0\;}{\ 0\;}$}{
        \ali{2.5em}{$\fxn{cmp}_0$} $\xleftarrow[]{128}
        \AVXArray{1..}{\ldots}{\ldots}{..1}$
    }
    \Else{
        \ali{2.5em}{$\fxn{cmp}_0$} $\xleftarrow[]{128}
        \AVXArray{0}{0}{0}{0}$
    }
    \BlankLine
    \If{$q =_{32}
    \AVXArray{\mathit{up}}{\mathit{up}}{\mathit{up}}{\mathit{up}}$}{
        \ali{2.5em}{$\fxn{cmp}_{\mathit{up}}$} $\xleftarrow[]{128}
        \AVXArray{1..}{\ldots}{\ldots}{..1}$
    }
    \Else{
        \ali{2.5em}{$\fxn{cmp}_{\mathit{up}}$} $\xleftarrow[]{128}
        \AVXArray{0}{0}{0}{0}$
    }

    \ali{2.6em}{$\fxn{test}_{0}$} $\xleftarrow[]{128} \fxn{cpm}_0 \bitwand
    \AVXArray{0}{5}{3}{1}$\\

    \ali{2.6em}{$\fxn{test}_{\mathit{up}}$} $\xleftarrow[]{128}
    \fxn{cpm}_{\mathit{up}} \bitwand \AVXArray{0}{6}{4}{2}$\\

    \ali{3em}{$r$}$\xleftarrow[]{128} (\fxn{test}_{0} \bitwor
    \fxn{test}_{\mathit{up}}) -_{32} \;\AVXArray{1}{1}{1}{1}$\\
    $f[0] \leftarrow r[3],\ f[1] \leftarrow r[2],\ f[2] \leftarrow r[1]$
\end{algorithm}%

\section{Test results for synthetic experiments}
\label{experiments}

In this section we present the results of the performance along with
the memory consumption of the algorithms, data structures and technologies
described at this work.
We compare standard, AVX and Morton ordered
quadrant implementations to each other using synthetic tests.

\subsection{Performance results.}

For testing purposes we create an array of 2396745 3D quadrants of
various refinement levels limited by a maximum of 7 and
call any algorithm to be measured in a loop over the quadrants.
We write the output of the algorithm to a local variable
to
prevent subsequent memory access.

We compile the code for the synthetic tests with the GNU GCC 10.3.0 compiler
using the compiler flag \texttt{-O3}, which enables automatic
vectorization, \texttt{-fno-math-errno}, \texttt{-DNDEBUG} and
\texttt{-DNVALGRIND}.

\begin{figure}
    \includegraphics[page=1, width=1\linewidth]{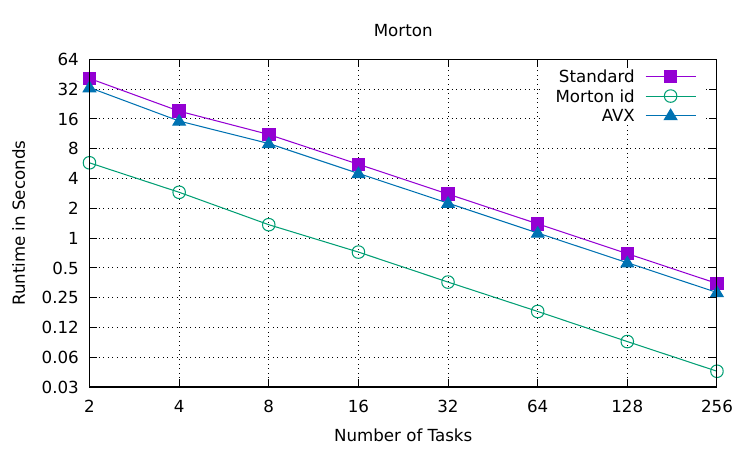}
    \caption{Strong scaling results for \fxn{Morton}
    (Algorithms~\ref{alg:standard_morton}, \ref{alg:morton_morton} and
    \ref{alg:avx_morton}) in various quadrant implementations.
    The raw Morton
    implementation gives a 77\% average performance boost and
    AVX2 gives 17\% in comparison to the standard quadrant implementation.}%
    \label{fig:synth_morton}%
\end{figure}
\begin{figure}
    \includegraphics[page=1, width=1\linewidth]{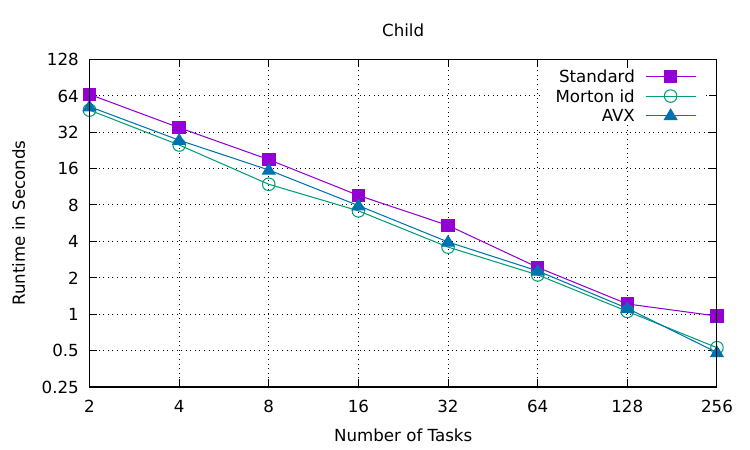}
    \caption{Strong scaling results for \fxn{Child}
    (Algorithms~\ref{alg:standard_child}, \ref{alg:morton_child} and
    \ref{alg:avx_child}) in various quadrant implementations.
    The raw Morton
    implementation gives 20\% while AVX2
    gives a 29\% average performance boost.}%
    \label{fig:synth_child}%
\end{figure}
\begin{figure}
    \includegraphics[page=1, width=1\linewidth]{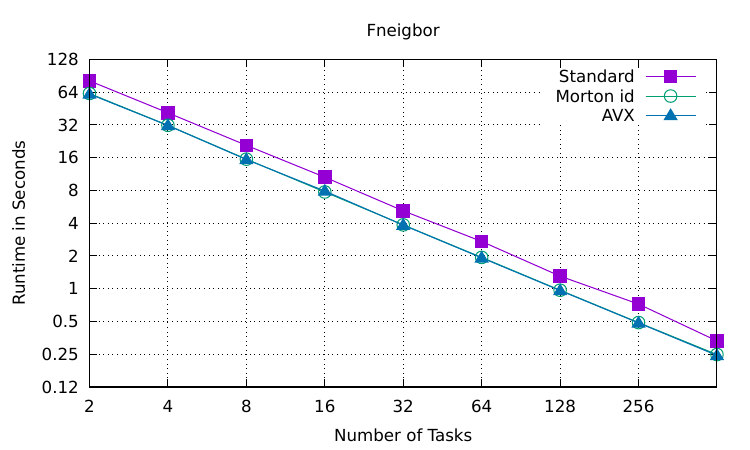}
    \caption{Strong scaling results for \fxn{FNeigh} (e.g.\
    Algorithm \ref{alg:face_neigh}) in various quadrant implementations.
    The raw Morton
    implementation gives 26\% while AVX2 gives a 27\%
    average performance boost.}%
    \label{fig:synth_fneigbor}%
\end{figure}
\begin{figure}
    \includegraphics[page=1, width=1\linewidth]{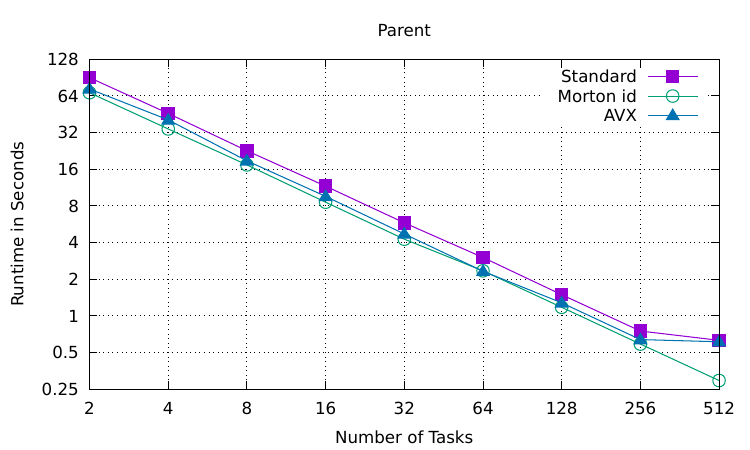}
    \caption{Strong scaling results for \fxn{Parent} (e.g.\
    Algorithms \ref{alg:morton_parent} and \ref{alg:avx_parent}) in
    various quadrant implementations.
    The raw Morton
    implementation gives
    27\% while AVX2 gives a 15\%
    average performance boost in comparison to the standard implementation.}%
    \label{fig:synth_parent}%
\end{figure}
\begin{figure}
    \includegraphics[page=1, width=1\linewidth]{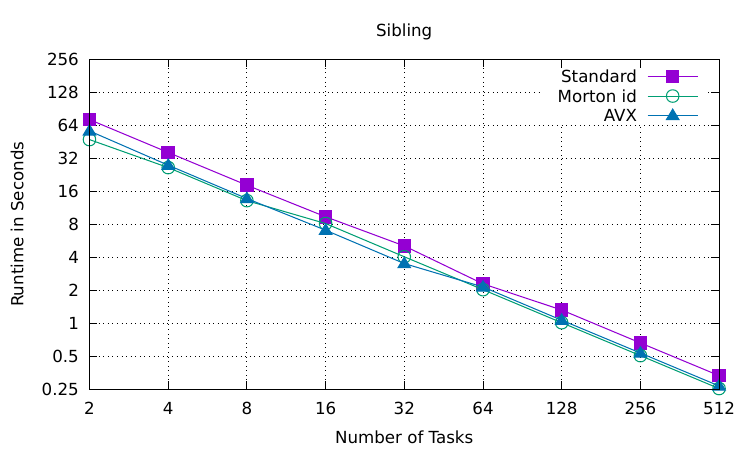}
    \caption{Strong scaling results for \fxn{Sibling} (e.g.\
    Algorithm \ref{alg:standard_sibling}) in various quadrant
    implementations.
    The raw Morton implementation gives 23\% while
    AVX2 gives a 21\% average
    performance boost.}%
    \label{fig:synth_sibling}%
\end{figure}
\begin{figure}
    \includegraphics[page=1, width=1\linewidth]{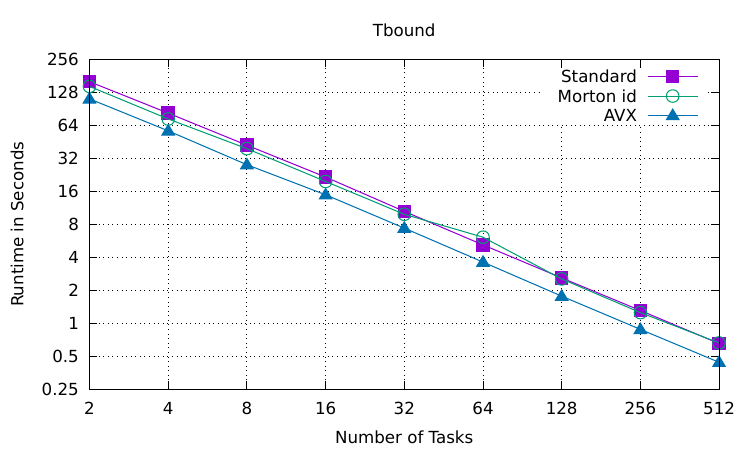}
    \caption{Strong scaling results for \fxn{Tree\_Boundaries}
    (i.e.\ Algorithm \ref{alg:avx_tree_bound}) in various quadrant
    implementations.
    The raw Morton implementation gives 3\% while
    AVX2 gives a 31\% average
    performance boost.}%
    \label{fig:synth_tbound}%
\end{figure}

Overall, both new quadrant implementations show speed-ups comparing to the
standard one.
The \fxn{Morton}, \fxn{Child} and \fxn{Parent} algorithms perform
faster for the raw Morton index than for AVX2 quadrants; see Figures
\ref{fig:synth_morton}, \ref{fig:synth_child} and \ref{fig:synth_parent}.
Conversely, \fxn{FNeighbor} and \fxn{Parent} make no
difference in terms of efficiency for both new representations (see Figures
\ref{fig:synth_fneigbor} and \ref{fig:synth_sibling}), while
\fxn{Tree\_Boundaries} works faster with quadrants processed with 128-bit
AVX2 instructions, which is demonstrated by Figure~\ref{fig:synth_tbound}.


\subsection{Comparison of memory consumption.}

We compare the amount of memory consumed by the program operating on the
various representations of quadrants.
Our software prescribes the following memory for each
quadrant implementation in 3D:
\begin{itemize}
    \item standard quadrants consisting of three 32-bit coordinates, 8 bits
    for level, 24 bits padding and 64 bits of user data.
    It consumes 24 bytes of memory in total.
    \item The raw Morton index representation as described in Section
    \ref{sec:mort} requires one 64-bit integer variable equal to 8 bytes of
    memory.
    \item Quadrants stored in extended 128-bit SSE/AVX2 registers as described
    in Section~\ref{sec:AVX} take 16 bytes.
\end{itemize}
We utilize a memory consumption preset of the Intel VTune Profiler software,
measuring the memory required to build a uniform octree of level 10 using
repeated calls to the Morton algorithm to obtain
the following results:
the program allocates 25.8 GB for standard, 17.2 GB for AVX and 8.6 GB for
Morton quadrants.  The ratio 3:2:1 between these values is as expected.

\section{Conclusion}

In this paper we describe two new representations of quadrants in the context of
the adaptive mesh refinement library \pforest, namely a single raw Morton
index long integer and a 128-bit SSE/AVX2 representation, respectively.  We
present the specific ways the quadrants are encoded for each representation
and detail several low-level algorithms operating on them.  Arguably, the
new algorithms are both less intuitive and more intricate than those based
on the standard representation using explicit $xyz$ coordinates.

Our tests demonstrate a performance boost over the standard for both new
types of implementation.
In most of them a raw Morton index shows better results than quadrants
processed by 128-bit SSE/AVX2 instructions, but in some cases the Morton
representation requires to make the algorithm more complex, which leads to
slowing down the performance.

Both new implementations require less RAM for
quadrant storage while keeping the maximum refinement level the same (18 for
the raw Morton index in 3D) or allowing it to be higher (31 for the SSE/AVX2
implementation).
We note however that the standard implementation has since caught up to a
maximum level of 29 in 3D, and its 8 bytes of payload might be removed for
a memory consumption on par with AVX2.
Based on these facts,
a user can choose their favorite implementation of a quadrant depending on
the required performance, resolution, and resources on the one hand and
simplicity and continuity on the other.

We consider several directions for the future.
The first one is the straightforward use of a wider register capacity,
for example 256-bit registers from AVX2 or 512 bits provided by AVX-512.
The practically reachable resolution can then be increased even further,
while the necessity of a refinement level beyond 30 is  admittedly somewhat
unclear.
The second is the integration of a raw Morton index implementation with
extended 128-bit CPU registers.
This will combine the advantages of both new implementations: lesser
complexity of some algorithms and higher maximum refinement level in
general.
Moreover, the use of one 128-bit register will still have equal RAM
requirements than AVX2 or the standard implementation without payload.

To make the new algorithms accessible through the \pforest software, we have
been working on a new branch of
high-level algorithms that operate on virtualized quadrants.
Presently we cannot predict whether the new interface and glue code will be
acceptable to the community, or if the way forward will rather be to wait
for a portable compiler support of 128 bit CPU registers and to hardcode the
appropriate updates into the internals of the software.


\section*{Acknowledgments}

This work is supported by a scholarship of the German Academic Exchange
Service (DAAD).

We acknowledge additional funding by the Bonn International Graduate School
for Mathematics (BIGS) as a part of the Hausdorff Center for Mathematics
(HCM) at the University of Bonn.
The HCM is funded by the German Research Foundation (DFG) under Germany's
excellence initiative EXC 59 -- 241002279 (Mathematics: Foundations, Models,
Applications).

We gratefully acknowledge partial support under DARPA Cooperative Agreement
HR00112120003 via a subcontract with Embry-Riddle Aeronautical University.
This work is approved for public release; distribution is unlimited.  The
information in this document does not necessarily reflect the position or
the policy of the US Government.

The authors gratefully acknowledge the granted access to the
Bonna compute cluster hosted by the University of Bonn.

\bibliographystyle{siam}
\bibliography{../bibtex/carsten,../bibtex/ccgo_new}
\end{document}

%% file: packages.tex
\usepackage{amsfonts,amsmath,amssymb}

\usepackage[algoruled, linesnumbered, vlined]{algorithm2e}
\usepackage{algorithmic}
\usepackage[small]{caption}
\usepackage{graphicx}
\usepackage{subfigure}
\usepackage{tikz}
\usepackage{url}
\usepackage{verbatim}
\usepackage{xcolor}
\usepackage{xspace}
\usepackage{multirow}

\usetikzlibrary{calc}

%% file: macros.tex
\newcommand{\algref}[1]{Algorithm~\ref{alg:#1}\xspace}

\newcommand{\dealii}{{\upshape deal.II}\xspace}
\newcommand{\petsc}{{\upshape PETSc}\xspace}
\newcommand{\pforest}{\texttt{\upshape p4est}\xspace}
\newcommand{\tetcode}{\texttt{\upshape t8code}\xspace}

\newcommand{\sZ}{\mathbb{Z}}

\newcommand{\cO}{\mathcal{O}}

\newcommand{\halfopen}[1]{[#1)}

\newcommand{\mitype}{\texttt{\_\_m128i}\xspace}
\newcommand{\mitypeex}{\texttt{\_\_m256i}\xspace}

\newcommand{\algrule}[1][.2pt]{\par\vskip.5\baselineskip\hrule height #1\par\vskip.5\baselineskip}

\newcommand{\ternary}[3]{(#1) \;?\; #2 \;:\; #3}
\newcommand{\bitwand}{\;\&\;}
\newcommand{\bitwneg}{\neg}
\newcommand{\bitwor}{\mathbin{|}}

\newcommand{\floors}[1]{\left\lfloor #1 \right\rfloor}
\newcommand{\ShiftLeft}{\ll}
\newcommand{\ShiftRight}{\gg}
\newcommand{\AVXArray}[4]{
\begin{array}{|c|c|c|c|}
	\hline
	\rule{0pt}{2.5ex} #1 & #2 & #3 & #4 \\ \hline
\end{array}
}
\newcommand{\AVXArrayRed}[2]{
    \begin{array}{|c|c|}
        \hline
        \rule{0pt}{2.5ex} #1 & #2 \\ \hline
    \end{array}
}
\newcommand{\ali}[2]{\makebox[#1][l]{#2}}

\newcommand{\fxn}[1]{\textup{\texttt{#1}}}
\newcommand{\vartype}[1]{\textup{\texttt{#1}}}

\SetNlSty{}{}{:}
\SetKwInOut{KwIn}{Input}
\SetKwInOut{KwResult}{Result}
\SetKwComment{Comm}{[}{]}
\SetCommentSty{commsty}
\SetAlgoNlRelativeSize{-1}